\documentstyle{article}
\input epsf
\begin{document}
\rightline{EFI-94-62}
\rightline{hep-th/9412074}
\vskip 1cm
\centerline{\LARGE \bf Spacelike Singularities and String Theory}
\vskip 1cm

\centerline{{\bf \Large Emil Martinec}}
\vskip .5cm
\centerline{\it Enrico Fermi Inst. and Dept. of Physics}
\centerline{\it University of Chicago,
5640 S. Ellis Ave., Chicago, IL 60637 USA}

\begin{abstract}
An interpretation of spacelike singularities in string
theory uses target space duality to relate the collapsing
Schwarzschild geometry near the singularity to an 
inflationary cosmology in dual variables.  
An appealing picture thus
results whereby gravitational collapse seeds the formation
of a new universe.  
%Observers in this daughter universe
%cannot look past their cosmological `singularity' because their
%effective lagrangian cannot describe it.
\end{abstract}
\vskip .5cm

{PACS numbers: 04.70.-s, 11.25.-w}
\vskip 1cm

%%%%%%%%%%%%%%%%%%%%%%%%%%%%%%%%%%%%%%%%%%%%%%%%%%%%%%%%%%%%%%%%%
\def\figloc#1#2{\vbox{{\epsfysize=3in
	\nopagebreak[3]
    \centerline{\epsfbox{fig#1.ps}}
	\nopagebreak[3]
    \centerline{Figure #1}
	\nopagebreak[3]
    {\raggedright\it \vbox{  #2 }}}}
    \bigskip
    }

\def\pref#1{(\ref{#1})}

\def\ie{{\it i.e.}}
\def\eg{{\it e.g.}}
\def\cf{{\it c.f.}}
\def\etal{{\it et.al.}}
\def\etc{{\it etc.}}

\def\inbar{\,\vrule height1.5ex width.4pt depth0pt}
\def\IC{\relax\hbox{$\inbar\kern-.3em{\rm C}$}}
\def\IR{\relax{\rm I\kern-.18em R}}
\def\A{{\bf A}}
\def\B{{\bf B}}
\def\C{{\bf C}}
\def\G{{\sst G}}
\def\T{{\bf T}}
\def\Z{{\bf Z}}
\def\One{{1\hskip -3pt {\rm l}}}
\def\sphere{{\bf S}^{\sst d-2}}

\def\sst{\scriptscriptstyle}
\def\tst#1{{\textstyle #1}}
\def\frac#1#2{{#1\over#2}}
\def\coeff#1#2{{\textstyle{#1\over #2}}}
\def\half{\frac12}
\def\hf{{\textstyle\half}}
\def\ket#1{|#1\rangle}
\def\bra#1{\langle#1|}
\def\vev#1{\langle#1\rangle}
\def\p{\partial}
\def\grav{{\sst \rm grav}}

%%%%%%%%%%%%%%%%%%%%%%%%%%%%%%%%%%%%
\def\np{{\it Nucl. Phys. }}
\def\pl{{\it Phys. Lett. }}
\def\pr{{\it Phys. Rev. }}
\def\ap{{\it Ann. Phys., NY }}
\def\prl{{\it Phys. Rev. Lett. }}
\def\mpl{{\it Mod. Phys. Lett. }}
\def\cmp{{\it Comm. Math. Phys. }}
\def\grg{{\it Gen. Rel. and Grav. }}
\def\cqg{{\it Class. Quant. Grav. }}
\def\ijmp{{\it Int. J. Mod. Phys. }}
\def\jmp{{\it J. Math. Phys. }}
\def\jetp{{\it Sov. Phys. JETP }} 
\def\jetplett{{\it JETP Lett. }} 
\def\dash{-----------------    }
\def\nextline{\hfil\break}
\catcode`\@=11
\def\slash#1{\mathord{\mathpalette\c@ncel{#1}}}
\overfullrule=0pt
\def\AA{{\cal A}}
\def\BB{{\cal B}}
\def\CC{{\cal C}}
\def\DD{{\cal D}}
\def\EE{{\cal E}}
\def\FF{{\cal F}}
\def\GG{{\cal G}}
\def\HH{{\cal H}}
\def\II{{\cal I}}
\def\JJ{{\cal J}}
\def\KK{{\cal K}}
\def\LL{{\cal L}}
\def\MM{{\cal M}}
\def\NN{{\cal N}}
\def\OO{{\cal O}}
\def\PP{{\cal P}}
\def\QQ{{\cal Q}}
\def\RR{{\cal R}}
\def\SS{{\cal S}}
\def\TT{{\cal T}}
\def\UU{{\cal U}}
\def\VV{{\cal V}}
\def\WW{{\cal W}}
\def\XX{{\cal X}}
\def\YY{{\cal Y}}
\def\ZZ{{\cal Z}}
\def\lam{\lambda}
\def\eps{\epsilon}
\def\vareps{\varepsilon}
\def\underrel#1\over#2{\mathrel{\mathop{\kern\z@#1}\limits_{#2}}}
\def\lapprox{{\underrel{\scriptstyle<}\over\sim}}
\def\lessapprox{{\buildrel{<}\over{\scriptstyle\sim}}}
\def\str{{str}}
\def\lstr{\ell_\str}
\def\gstr{g_\str}
\def\Mstr{M_\str}
\def\lpl{\ell_{pl}}
\def\Mpl{M_{pl}}
\def\endpt{{\sst endpt}}
\def\max{{\sst\rm max}}
\catcode`\@=12

\hyphenation{in-fla-tion-ary}
%%%%%%%%%%%%%%%%%%%%%%%%%%%%%%%%%%%%%%%%%%%%%

String theory is strongly believed to solve the short-distance problems
of quantum gravity by providing a fundamental length scale 
$\lstr=\sqrt{\hbar c/T}$, where $T$ is the string tension.
Perturbative studies of high-energy string scattering 
(Gross and Mende 1988, Amati \etal\ 1988), 
the stringy resolution of orbifold singularities
(Dixon \etal\ 1985),
and target-space duality (for a review, see Giveon \etal\ 1994) 
all point in this direction.
Thus one might expect that there is no such thing as a `naked' 
or timelike singularity in string theory; the exponentially soft
high energy behavior of the theory (in weak coupling, at least)
will forbid any observation of the kinds of arbitrarily violent processes
associated with high field gradients over small regions.
The decay of strong electromagnetic fields by string pair
production (Bachas and Porrati 1992; see also Fradkin and Tseytlin 1985,
Russo and Tseytlin 1994)
also supports this picture of the string's aversion to large field
strengths.
Spacelike singularities, such as those encountered in 
black hole formation, seem to mount a stronger challenge.  
In this case, the singularity is in the future; all the soft
high-energy behavior is helpless in the face of the crushing
together of future-directed light cones inside the black hole horizon.
No causal force can prevent collapse; indeed the soft high-energy
behavior of strings at high energy would seem to act
in the wrong direction, placing an upper
limit on any potential stabilizing force.
How then can string theory be a complete theory of nature if it
does not tell us how to evolve geometry through a spacelike
singularity?  I will argue here that this question 
presupposes an improper treatment of string geometry near
the singularity.  

Target-space duality (often called T-duality)
appears to be a pervasive feature of
classical solutions in string theory (\cf\ Giveon \etal\ 1994); 
it has been demonstrated
in toroidal and orbifold geometries (Kikkawa and Yamasaki 1984,
Sakai and Senda 1986, Nair \etal\ 1987), cosmological settings 
(\cf\ Tseytlin and Vafa 1992, Veneziano 1991),
and even the moduli space of $K3$ surfaces (Aspinwall and
Morrison 1994; see also Seiberg 1988).  
The use of mirror symmetry in the last case
indicates a similar duality of Calabi-Yau threefolds.
In all these instances, no regime exists
in which the proper low-energy interpretation of the
geometry involves a manifold with a scale size $R\ll\lstr$.
Rather, as the geometry is taken below scale sizes of order $\lstr$,
the low-energy effective lagrangian typically breaks down
via the appearance of new soft modes
(\eg\ winding modes in toroidal and orbifold geometries);
simultaneously, the modes of the original low-energy fields
become stiff.
Decreasing the scale size further, the original low-energy fields
have momenta larger than $1/\lstr$ and should no longer 
be included in the effective action, whereas the new soft modes
become even softer and comprise the fields with which 
low-energy dynamics is properly understood.  
This new dynamics can be dramatically different 
from that of the original
effective field theory; it can change the low-energy field content
(Dine \etal\ 1989), gauge group (Narain \etal\ 1987, Ginsparg 1987),
and even the spacetime topology (Kiritsis and Kounnas 1994).

With this characteristic of string geometry in mind, let us
examine the Schwarzschild solution (the analysis applies without
essential modification to cosmological collapse situations).  
Suppose that string
theory is weakly coupled: $\Mpl=\Mstr/\gstr$
with the string coupling
$e^{-\Phi}=\gstr\ll 1$; and consider large black holes $M\gg \Mstr$
(the Schwarschild radius $R=2 g^2 \lstr^2 M\gg \lstr$).
The line element is
\begin{equation}
  ds^2=-\bigl[1-\bigl(\coeff{2M}{{\textstyle r}}\bigr)^{d-3}\bigr]dt^2 +
	\bigl[1-\bigl(\coeff{2M}{{\textstyle r}}\bigr)^{d-3}\bigr]^{-1}
		dr^2 
	+r^2 d\Omega_{{\sst d-2}}^2\ .
\end{equation}
Inside the horizon, $t$ is a spacelike coordinate
whereas $r$ is timelike.  Therefore let us define 
$\tau=[2R/(d-1)](r/R)^{(d-1)/2}$
and $\rho=t$, so that near the singularity
\begin{eqnarray}
  ds^2&\sim& -d\tau^2 + a(\tau)\,d\rho^2+b(\tau)\,
		d\Omega_{{\sst d-2}}^2\nonumber\\
	&=&-d\tau^2 + 
	\bigl[\coeff{(d-1)\tst\tau}{2 R}\bigr]^{-{2(d-3)}/{(d-1)}}
		d\rho^2
	+R^2\bigl[\coeff{(d-1)\tst\tau}{2R}\bigr]^{4/{(d-1)}}
		d\Omega_{{\sst d-2}}^2\ .
\label{KS_metric}
\end{eqnarray}
One can regard this metric as an anisotropic homogeneous
cosmological model with spatial sections 
of topology $\IR\times\sphere$ (\cf\ MacCallum 1979 for a review).  
As the singularity approaches, the scale factor of the $\IR$ direction
stretches to infinity while the radius of the $\sphere$
shrinks to zero.  However, at $\tau\sim\lstr$ -- 
a thickened hypersurface
we shall call the `threshold' for lack of
a better term\footnote{Perhaps `stretched singularity' would
conform more to current usage; 
purgatory ({\sl `a place or state of temporary suffering or misery'} -- 
Webster's ninth New Collegiate Dictionary) might also be an apt
description of this region.} -- we should
switch the description to a dual effective action
of the $\sphere$.  {\sl Very} crudely speaking, we should take 
\begin{equation}
	b(\tau)\rightarrow 1/b(\tau)
\label{metric_transf}
\end{equation}
while shifting the string coupling
\begin{equation}
	\Phi\rightarrow\Phi-(d-2)\log b\ .
\label{dil_transf}
\end{equation}
In this regime the kinetic terms (the extrinsic
curvature of spatial sections) dominate over potential terms
(intrinsic spatial curvature) in the Hamiltonian constraint equation;
hence to a first approximation we can perhaps ignore the curvature of
the sphere and borrow the familiar duality transformation properties 
\pref{metric_transf},\pref{dil_transf} of flat (toroidal) geometries.
Of course, large field gradients will strongly modify the
specific form of the duality transformation.
Our strategy will be to build confidence in situations where we can
make reliable perturbative expansions, and then extrapolate
to the regime of rapidly varying scale factors.

For spatial topology $\IR\times({\bf S}^1)^{(d-2)}$
carrying metrics of the form
\begin{equation}
	ds^2=-dt^2+\sum_i(a_i(t)dx_i)^2\ ,
\label{kasner}
\end{equation} 
equations \pref{metric_transf},\pref{dil_transf} are indeed
the correct duality transformations for slowly evolving radii $a_i$
(Veneziano 1991).
If $a_1$ (the $\IR$ scale factor)
is large and increasing while $a_2,\ldots,a_{d-1}$
decrease from $a>\lstr$ to $a<\lstr$, we would have no qualms
about changing the low energy fields from momentum to winding modes
at $a\sim\lstr$.  We can make this evolution adiabatic by allowing the
dilaton field $\Phi$ to evolve in time (see \eg\ Tseytlin 1992c,
Kiritsis and Kounnas 1994).
Since the target space duality symmetry in the
toroidal case is due to an easily
identifiable structure (symmetry between momentum and
winding number), we can plausibly expect duality to persist
even during rapid changes of the radii -- for instance near
the Kasner-type singularity (\cf\ MacCallum 1979)
\begin{equation}
  ds^2= -\lstr^2d\tau^2 + \tau^{-2(d-3)/(d-1)}dx_1^2
	+ \tau^{4/(d-1)}\sum_{i=2}^{d-1} dx_i^2
\end{equation}
analogous to \pref{KS_metric}.
Thus for $\tau\lessapprox 1$ we should
switch to the dual theory \pref{metric_transf},\pref{dil_transf}:
\begin{eqnarray}
	\widetilde{ds}^2&\sim  &
		-\lstr^2d\tau^2 + \tau^{-2(d-3)/(d-1)}dx_1^2
		+\tau^{-4/(d-1)}\sum_{i=2}^{d-1} dx_i^2\nonumber\\
	\Phi&\sim& \log g_{\str,0}-
		2\bigl(\coeff{\tst{d-2}}{\tst{d-1}}\bigr)\log\tau\ .
\label{Kasner_metric}
\end{eqnarray}
This metric describes a so-called super-inflationary universe
(inflation to infinite scale factor in finite comoving time
(Lucchin and Matarese 1985)).
The inflation is driven by the kinetic energy of the dilaton
field, which is running toward strong coupling.

Exact vacuum solutions of string theory may be found in the similar
situation of gravitational
plane wave solutions (Horowitz and Steif 1990,
Tseytlin 1992b, Polchinski and Smith 1991), 
where $a_i$ evolve as a function
of a null coordinate $u$.  One can choose the wave profile
$a_i(u)$ to evolve between any given initial and final values as
the wave passes; in particular there is no restriction
on the gradient of the scale factor. 
The string coupling can be made to
change by a finite amount across the wave,
hence the theory can be made arbitrarily weakly coupled.
One can even contemplate a null singularity, where the
metric blows up at finite $u$ (Horowitz and Steif 1990, 
de Vega and Sanchez 1992) -- a kind of null version of the
Schwarzschild singularity.  An analysis of this situation is in progress.

Generically, a dilaton rolling toward weak coupling
provides a friction that damps the growth of the scale factor,
while evolution to strong coupling has a runaway behaviour.
The minisuperspace cosmological equations read
(\cf\ Tseytlin and Vafa 1992, Tseytlin 1992a, Veneziano 1991)
\begin{eqnarray}
{\dot\varphi}^2-\sum_{i=1}^{d-1}{\dot\lam}_i^2
		&=&2U + 2 e^\varphi E\nonumber\\
	{\ddot\lam}_i-\dot\varphi{\dot\lam}_i
		&=&-\frac{\p U}{\p \lam_i}+e^\varphi P_i\nonumber\\
	\ddot\varphi-\sum_{i=1}^{d-1}{\dot\lam}_i^2
		&=&\frac{\p U}{\p\varphi}+e^\varphi E
\label{minisuperspace}
\end{eqnarray}
in terms of the shifted dilaton $\varphi=2\Phi-\sum_i\lam_i$
and the scale factor $\lam_i=\log(a_i)$; $U(\varphi,\lam_i)$ is the
effective potential and $E$, $P_i$ are the energy and pressure
of a perfect fluid of string matter.  One easily sees from
these equations that $\dot\varphi<0$ damps the expansion
of spatial volume, which then freezes the dilaton.
On the other hand, $\dot\varphi>0$ is unstable to growing
spatial volume, which then accelerates the growth of
the dilaton, \etc, generating super-inflation.
Depending on the form of matter stress-energy and its equation
of state, one can obtain a number of different behaviors
including oscillatory scale factors (Tseytlin 1992c).
An appealing scenario has an initial era of dilaton-driven inflation;
then the dilaton freezes or
settles toward a weak-coupling regime while string matter
generated during inflation
acts as a source for further cosmological expansion.
Strings in a (super)inflationary universe are unstable 
in the sense that the
string size grows as the scale factor (Gasperini \etal\ 1991).  
If the inflationary
era terminates, the produced strings might serve as a source
of further cosmological expansion (\cf\ Turok 1988, Barrow 1988) 
as well as the seeds for structure formation.
I must emphasize that the T-dual effective lagrangian
will not be quantitatively valid in the regime we wish to use it,
since the metric and dilaton are changing rapidly 
on scales of order $\lstr$.  Nevertheless, we may hope to borrow
intuition gained from model situations (like the plane wave solutions)
where one can continuously vary the geometry between gentle
and violent evolution of the scale factors.

Unfortunately, at present one can only speculate on the
mechanism, if any, that terminates this dilaton-driven inflation
(for a discussion of the problems, see Brustein and Veneziano 1994).  
Perhaps, at sufficient dilaton velocity, higher string corrections
to the equations of motion
absorb the dilaton kinetic energy driving the expansion,
and turn the dilaton evolution back toward weak coupling.
Alternatively, if strong-weak coupling duality (so-called S-duality)
is a feature of string theory, once the inner, T-dual universe
inflates to the strong coupling regime, we change effective
descriptions again so that the theory is headed toward weak
coupling in the ST-dual theory with interaction strength $1/\gstr$
(\cf\ Horne and Moore 1994).
It must be emphasized that evidence for strong-weak coupling
duality in string theory is meager at best.  Any quantity
receiving perturbative corrections cannot be tested without
an understanding of nonperturbative string dynamics (\cf\
Sen 1994).  Hence one can as yet investigate only
renormalizable gauge theories having enough
supersymmetries that classical relations are perturbatively
exact for the quantity of interest
(Sen 1994, Seiberg and Witten 1994; 
for intriguing hints of what may lie beyond,
see Gauntlett and Harvey 1994).  One thus cannot be sure whether
S-duality is a property of string theory or merely of this
restricted class of gauge theories.  In this context, I should
point out that the form of the low-energy effective action
which manifests S-duality is written using the Einstein metric 
$G^E_{ab}$ for which it is natural to think of $\Mpl=\Mstr e^{-\Phi}$
as fixed.  However for string theory it is more natural
to use the sigma-model metric $G^\sigma_{ab}=e^{\alpha\Phi}G^E_{ab}$
in which the fundamental constant
$\Mstr$ is fixed.  Hence the S-dual of a 
string theory with $\Mpl=\Mstr e^{-\Phi}$ will be one with
$\Mpl=\Mstr e^{+\Phi}$.  Curiously the latter is precisely the relation
between the string and Planck scales in open string theory
if the former is that of heterotic strings
(Dine and Seiberg 1985; see also Polchinski 1994), suggesting
that S-duality might relate rather different sorts of
weakly coupled string theories if it holds.

We expect phenomena quite similar to the toroidal example
to occur in the 
$\IR\times\sphere$ topology relevant to gravitational
collapse.  The main difference lies in the fact that T-duality
is a more complicated transformation in this case, mixing all
the string modes even for an adiabatically changing scale factor.
There is no clean separation between momentum and `winding' modes.
Nevertheless, we expect soft modes arising from the fact that entropy of
string wandering is less suppressed than center of mass momentum
at scale factors much smaller than $\lstr$.
By maintaining a wide separation of scales
$\lstr\gg\lpl$, we expect that the transition
to the dual geometry occurs before quantum gravity
effects become important.
This is not to say that quantum effects will be negligible
before the coupling grows strong.
Curvatures and field strengths are of order the string scale
in the transition region and in the dual geometry, 
so string creation will have an important effect on the subsequent 
expansion in the T-dual theory.
For instance, Zel'dovich (1970)
%(Zel'dovich 1970) 
has argued that particle creation rapidly isotropizes
the expansion of initially anisotropic cosmological models
such as \pref{Kasner_metric}.  
One can imagine that the aversion to high field strength 
of the sort exhibited
in the situation of constant electromagnetic fields 
(Bachas and Porrati 1992, Russo and Tseytlin 1994) 
could provide a drag on the expansion
through copious string production.  
The production in a classical background
is $O(\hbar)$ but independent of $\gstr$
at lowest order, so it can be large when field strengths are $O(1)$
in natural string units even when $\gstr$ is small.
In the electromagnetic field example, magnetic fields
in excess of a critical value cause a number of
string modes to become light and then tachyonic;
critical electric fields cause the pair production rate
for an infinite number of modes to diverge.
The back reaction of the produced pairs rapidly acts to dissipate the
field.  In the gravitational case string production should provide
energy density to slow the expansion to a rate of order one
in string units.  When the expansion rate exceeds one in natural
string units, tidal forces exceed the string tension and any string
grows as the scale factor.  There is no obvious quantum number
that prevents spontaneous tree-level
generation of strings -- for instance zero momentum dilatons --
which then start to grow without bound.  The produced strings'
energy density will act to slow expansion.

In fact, curvature becomes  
large already in the Schwarzschild region at a $(d-2)$-sphere radius
of order $r_0\sim R(\lstr/R)^{2/(d-1)}\gg\lstr$; thus one might expect
strong quantum effects {\it before} the $(d-2)$-spheres 
become small.  Several groups (Israel and Poisson 1988, 
Frolov \etal\ (1990)) have speculated that strong
quantum particle production causes a transition to an inflationary
de Sitter phase, thereby avoiding 
gravitational collapse in a manner quite similar to the present
proposal but without resorting to string theory
(although since their scenario is particle-theoretic,
the subsequent `inner' universe is made out of ordinary matter rather
than stringy dual matter; also $\lstr$ in our considerations is
replaced by $\lpl$ in theirs).
%It is doubtful that quantum effects avert gravitational collapse
%entirely.  
However, the elapsed proper time between spheres of radius $r_0$ and
radius $\lstr$ is the string scale $\lstr/c$.  To realize this
de Sitter transition, quantum effects
would need to {\it reverse} the momentum of the geometry (the extrinsic 
curvature of spatial hypersurfaces) over an essentially meaningless
time scale of order one in string units; whereas all that need
be done to control the curvature is to {\it slow the rate of change}
of the momentum.
Whether quantum effects avert, enhance or merely delay 
gravitational collapse remains to be seen, and will require a 
better understanding of string geometry at large field gradients.
String theory offers the possibility that the collapse need not be
stopped, with field momentum built up during collapse
used to fuel expansion into
a large universe using new degrees of freedom unavailable 
within the realm of field theory.

Let us now extend the scope of our speculations
to the issue of black hole evaporation (Hawking 1975).
We only expect the metric \pref{KS_metric} to apply outside
the collapsing matter and until the endpoint of the Hawking
evaporation process, \ie\ in an interval 
$\rho({\it formation})<\rho<\rho({\it endpoint})$.
At $\rho(\it endpt)$, an accelerating accumulation 
of negative stress-energy 
-- heuristically the partners of the radiated Hawking particles --
strikes the `threshold', which might then turn timelike.
The spacelike transition region to the dual cosmology is bracketed
by two implosive events; the nearly null collapse of infalling
matter, and the final flash of radiation at the endpoint.
What could the geometry look like afterward?
The original spacetime will be Schwarzschild (together with
outgoing Hawking radiation) down to a radius of order $\lstr$.
The cosmology of the dual universe beyond the threshold will
presumably continue to evolve.
There will then presumably be a spacelike path of stringy dimensions
through a wormhole connecting the two regions.  
It is unlikely that the threshold could pinch off 
and disconnect the two worlds; this would require precisely the
kind of singularity that we have argued should not occur in string theory.
The geometry has features of a stable remnant from the viewpoint
of the original (Schwarzschild) universe, as well as
of baby universe formation; both objects have been considered in
the context of the evaporation process (Aharonov \etal\ 1987,
Banks \etal\ 1992; Zel'dovich 1977, Dyson 1976).
For a crude picture of the
geometry, see figure 1
(it is assumed that inflation of the dual geometry
terminates at some point).  
Because of the large volume of the dual geometry beyond
the (now timelike) threshold, this sort of remnant should be
hard to pair produce (Banks \etal\ 1993).

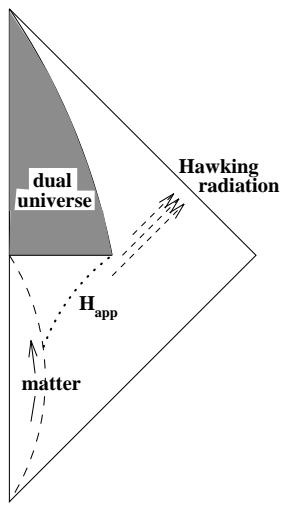
\figloc1{Penrose diagram for 
a black hole which forms and subsequently evaporates.  
The dashed line indicates the trajectory
of collapsing matter; the dotted line the apparent horizon.}

One can again imagine
adiabatic transition regions that could serve as a model
for how the geometry might appear
well after the endpoint.  Consider again
the toroidal universe \pref{Kasner_metric}, but now allow
the torus radii $\lam_i=\log(a_i/\lstr)$ to vary adiabatically 
in the noncompact $x_1$ direction about a central value of
zero (\ie\ $a_i\sim\lstr$).  
For instance we can consider a `standing wave' 
$\lam_2=\log(R_\max)\,\cos(x_1/L)\cos(t/L)$ (see figure 2).
This will not satisfy the string equations of motion,
but we may compensate by allowing adiabatic variation of
some of the other fields (similar solutions are described
in Kiritsis and Kounnas (1994)).  
An observer at $x_1=0$ will see a large $(d-1)$-dimensional 
spatial universe; generic objects in his world carry momentum in the
$x_2$ direction, which is nearly continuous.  
However the gap in $p_2$ gets large near $x_1=\pi/2$ and 
the spatial world looks effectively
$(d-2)$-dimensional with a spatially varying mass
$M_{d-2}=p_2\,\exp[-\lam_2(x_1,t)/\lstr]$.  
From the $(d-1)$-dimensional
viewpoint the geometry is conical in this region; there is
an `angular momentum barrier' to probing the tip of the cone.  
A test particle carrying nonvanishing
$p_2$ sent to probe the world at $x_1=L\pi/2$ will be reflected
by this effective potential.  It cannot penetrate into the
dual world because it is a macroscopic winding string there
with energy of order $R_\max/\lstr^2$.  The observer's world is
effectively of finite extent $L\pi$ in the $x_1$ direction.
He sees a `big bang' in his past at $t=-L\pi/2$ 
and will experience a `big crunch' at $t=L\pi/2$ when he will be
stretched on the cosmic rack of the dual world that follows.
Observers in this universe
cannot look past their cosmological `singularity' because their
effective lagrangian cannot describe it.
Only the nongeneric modes with $p_2=0$ can pass into the dual world,
like the S-waves of GUT monopoles (Callan 1982; Rubakov 1981).
One might expect a `threshold' remnant to have similar properties:
only very special modes can pass into or out of the dual world, and
for most purposes it would look to the Schwarzschild observer
as a soliton of mass $O(\Mstr)$.
\figloc2{Toroidal model of a remnant; a transition region
forms a `domain wall' between two regions of spacetime
described by T-dual geometries.  Reading from left to right gives a
picture of the evolution in $t$ for fixed $x_1$. }

Unfortunately, there is no obvious barrier to bidirectional
energy flow across a timelike threshold.  What prevents
the hot dual string gas beyond the threshold from leaking
into our universe, violating energy conservation?  The answer
may require good control over the geometry in the threshold
region to see if a barrier might arise; perhaps the neutral
tubular `remnant' described by Giddings \etal\ (1993)
could provide a useful starting point.  
Alternatively, the problem can be avoided altogether if the threshold
always lies to the future of the exterior Schwarzschild region.
This occurs when the threshold asymptotes to a null curve
(figure 3) as 
the evaporation comes to an end.\footnote{A possibility suggested to
me by M. O'Loughlin.} 
\nopagebreak[3]
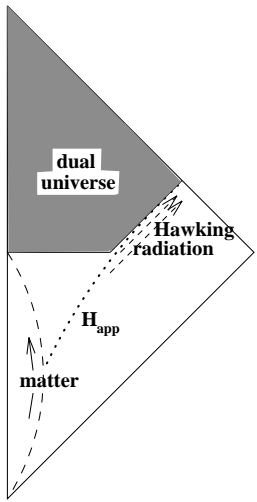
\figloc3{Alternative scenario for black hole evaporation, 
where the threshold asymptotes to a null trajectory.}

We have explored a T-duality model of collapse and 
subsequent re-expansion, but there
may well be other models. In the algebraic geometry of Calabi-Yau
sigma models, there are paths in the K\"ahler
moduli space (Witten 1993, Aspinwall \etal\ 1994)
along which a homological two-sphere shrinks to `zero' size, then
expands to positive radius in another geometry.  From the
viewpoint of the original low-energy description, the 
two sphere evolves to {\it negative} radius.
The dilaton expectation value remains constant along the path.
In gravitational collapse, 
it may be that the `zero radius' boundary in the space of
metrics on the collapsing $\sphere$ joins smoothly onto another
geometry in which it is expanding.  The key feature to both this model
and the T-duality model of collapse is the re-expansion
toward large spatial volume in a different semi-classical geometry.

The property of string theory that we want to exploit
is the absence of geometries with scale size smaller than $\lstr$.
Singularity of the geometry is avoided simply if the collapsing
spheres never reach a singular state; although perhaps somewhat
more aesthetically pleasing from a cosmological standpoint, 
re-expansion is not essential.
The $\sphere$ could  stabilize at some small
radius of order $\lstr$ as in the `cornucopion' models of
Banks \etal\ (1992).  If we regard $\tau$ in \pref{KS_metric}
as renormalization group time, the results of Cecotti and Vafa (1992)
point toward the stabilization of the geometry of N=2 supersymmetric 
sigma models as they continue through `zero radius';
however in the ${\bf CP^n}$ models they studied in detail,
it is not apparent whether the geometry reexpands.
If it does not, one would then need to explain how 
the kinetic energy of the collapsing geometry is dissipated.

To summarize, target space duality (or some similar model of string
geometry) may offer a means to discover
the manner in which string theory resolves spacelike singularities.
At the very least, it indicates that all issues of singularities
in string theory might be recast as questions of (perhaps
strong coupling) dynamics
in large spatial volumes rather than problems of infinite field strength 
and consequent breakdown of dynamics.

\vskip 1cm
\leftline{\bf \large \noindent Acknowlegements}
\vskip .5cm

Thanks to 
T. Banks,
D. Kutasov,
R. Wald,
and especially
M. O'Loughlin
for helpful discussions.

\vskip 1cm
\centerline{\bf Note added}
\vskip .5cm
After this work was completed, E. Kiritsis and C. Kounnas informed
me that they had also proposed target space duality as a means
of resolving black hole singularities (\cf\ Kiritsis 1993).  I would
like to thank them for bringing this work to my attention.

\vskip 1cm
\leftline{\bf \large \noindent References}
\vskip .5cm
\begin{itemize}
\itemindent=-18pt\itemsep=0pt\parsep=0pt
\item[] Aharanov Y, Casher A and Nussinov S 1987 \pl {\bf 191B} 51 
%remnants
\item[] Amati D, Ciafaloni M and Veneziano G 1988 \ijmp {\bf A3} 1615
%high en scatt
\item[] Aspinwall P, Morrison D 1994 {\it String Theory on K3 Surfaces}, 
Duke Univ. preprint DUK-TH-94-68, hep-th/9404151 
%K3 duality
\item[] Aspinwall P, Greene B and Morrison D 1994 \np {\bf B416} 414
%CY mod sp, mirror mflds and spacetime top ch in string theory
\item[] Bachas C and Porrati M 1992 \pl {\bf 296B} 77
%string production in E field
\item[] Banks T, Dabholkar A, Douglas M and O'Loughlin M 1992 
\pr {\bf D45} 3607 
%cornucopions
\item[] Banks T, O'Loughlin M and Strominger A 1993 \pr {\bf D47} 4476
%remnant pair prod
\item[] Barrow J D 1988 \np {\bf B310} 743 
%string inflation
\item[] Brustein R and Veneziano G 1994 \pl {\bf 329B} 429
%graceful exit problem in string cosmology 
\item[] Callan C 1982 \np {\bf B212} 391 
%monopole catalysis
\item[] Cecotti S and Vafa C 1992 \prl {\bf 68} 903
%exact results for supersymmetric sigma models 
\item[] de Vega H and Sanchez N 1992 \pr {\bf D45} 2783 
%strings in grav waves
\item[] Dine M and Seiberg N 1985 \prl {\bf 55} 366 
%couplings and scales in string theory
\item[] Dine M, Huet P and Seiberg N 1989 \np {\bf B322} 301
%large and small radius in string theory
\item[] Dixon L, Harvey J, Vafa C and Witten E 1985 \np {\bf B261} 678
%strings on orbifolds
\item[] Dyson F 1976 IAS preprint, unpublished 
%baby universes
\item[] Fradkin E S and Tseytlin A A 1985 \pl {\bf 163B} 123
%open strings in E field; 
%also Tseytlin 1986 \np B276 391 ,ERRATUM-ibid.B291:876,1987
\item[] Frolov V P, Markov M A and Mukhanov V F 1990 \pr {\bf D41} 383
%black holes as possible sources of closed universes
\item[] Gasperini M, Sanchez N and Veneziano G 1991 
\ijmp {\bf A6} 3853 
%unstable strings in inflationary cosmology
\item[] Gauntlett J and Harvey J 1994 {\it S Duality and 
the Spectrum of Magnetic Monopoles 
in Heterotic String Theory}, 
U. Chicago preprint EFI-94-36 hep-th/9407111 
%H monopoles and S-duality
\item[] Giddings S, Polchinski J and Strominger A 1993 \pr {\bf D48} 5784
%4d black holes in string theory
\item[] Ginsparg P 1987 \pr {\bf D35} 648
%toroidal moduli space
\item[] Giveon A, Porrati M and Rabinovici E 1994 
{\it Phys. Rep.} {\bf 244C} 77 
%duality review
\item[] Gross D and Mende P 1988 \np {\bf B303} 407 
%high en scatt
\item[] Hawking S W 1975 \cmp {\bf 43} 199 
%BH evaporation
\item[] Horne J and Moore G 1994 {\it Chaotic Coupling Constants},
Yale Univ. 
preprint YCTP-P2-94, hep-th/9403058 
%chaotic coupling consts
\item[] Horowitz G and Steif A 1990 \pr {\bf D42} 1950
%plane waves
\item[] Israel W and Poisson E  1988 \cqg {\bf 5} L201
%structure of the black hole nucleus
\item[] Kikkawa K and Yamasaki M 1984 \pl {\bf 149B} 357 
%torus duality
\item[] E Kiritsis 1993
{\it Duality Symmetries and Topology Change in String Theory},
International Europhysics Conference on High Energy Physics 
(Paris: Editions Fronti\`eres)
%idea on spacelike singularities; hep-th/9309064
\item[] Kiritsis E and Kounnas K 1994 \pl {\bf 331B} 51
%topology change

\item[] Lucchin F and Matarese S 1985 \pl {\it 164B} 282 
%super-inflation
%\item[] Lukash V N and Starobinsky A A 1974 \jetp {\bf 39} 742
%back reaction in cosmology
\item[] MacCallum M A H 1979 {\it General Relativity:
An Einstein Centenary Survey} 
(Cambridge: Cambridge University Press) p533
\item[] Nair V P, Shapere A, Strominger A and Wilczek F 1987 
\np {\bf B287} 402 
%torus duality
\item[] Narain K S, Sarmadi M H and Witten E 1987 \np {\bf B279} 369
%toroidal moduli space
\item[] Polchinski J 1994 {\it Combinatorics of Boundaries 
in String Theory},
ITP preprint NSF-ITP-94-73 hep-th/9407031
%nonpert effects and open strings
\item[] Polchinski J and Smith E 1991 \pl {\bf 263B} 59 
%duality survives time dependence
\item[] Rubakov V 1981 {\it JETP Lett.} {\bf 33} 644
%monopole catalysis
\item[] Russo J G and Tseytlin A A 1994 
{\it Constant magnetic field in closed string theory: 
An Exactly solvable model}, CERN preprint CERN-TH-7494-94,
hep-th/9411099
\item[] Sakai N and Senda I 1986 {\it Prog. Theor. Phys.} {\bf 75} 692
%torus duality
\item[] Seiberg N 1988 \np {\bf B303} 286 
%moduli space of CY
\item[] Seiberg N and Witten E 1994 \np {\bf B426} 19 
%S duality in N=2YM
\item[] Sen A 1994 \ijmp {\bf A9} 3707 
%S duality review
\item[] Tseytlin A A 1992a  
{\it String Quantum Gravity and Physics at the Planck Energy
   Scale: Erice proceedings} (Singapore: World Scientific) p202
%string cosmology and dilaton (Erice Theor. Phys. 1992:202-223)
\item[] \dash 1992b \pl {\bf 288} 279 
%null sigma models
\item[] \dash 1992c \cqg {\bf 9} 979 
%dilaton, winding modes & cosm solns
%\item[] \dash 1993 \np {\bf B390} 153 
%null sigma models
\item[] Tseytlin A A and Vafa C 1992 \np {\bf B372} 443
%elements of string cosmology 
\item[] Turok N 1988 \prl {\bf 60} 549
%string driven inflation
\item[] Veneziano G 1991 \pl {\bf 265B} 287 
%scale factor duality in cosm
\item[] Witten E 1993 \np {\bf B403} 159
%phases of N=2 sigma models
\item[] Zel'dovich Ya B 1970 \jetplett {\bf 12} 307
%particle prod in anisotropic cosm
\item[] \dash 1977 \jetp {\bf 45} 9 
%baby universes
%\item[] Zel'dovich Ya B and Starobinsky A A 1972
%{\it Sov. Phys.: JETP} {\bf 34} 1159 
%particle prod in anisotropic cosm

\end{itemize}

\end{document}